\documentclass[conference]{IEEEtran}

\usepackage{cite}
\usepackage{amsmath,amssymb,amsfonts}
\usepackage{algorithmic}
\usepackage{graphicx}
\usepackage{textcomp}
\usepackage{xcolor}
\usepackage{booktabs}
\usepackage{multirow}
\usepackage{url}
\def\BibTeX{{\rm B\kern-.05em{\sc i\kern-.025em b}\kern-.08em
    T\kern-.1667em\lower.7ex\hbox{E}\kern-.125emX}}

\begin{document}

\title{Optimising Neural Speech Codecs for 300 bps Communication Using Reinforcement Learning}

\author{
\IEEEauthorblockN{
Junyi Wang\textsuperscript{1,*},
Chi Zhang\textsuperscript{1,*},
Jing Qian\textsuperscript{2},
Haifeng Luo\textsuperscript{2},
Hao Wang\textsuperscript{2},
Zengrui Jin\textsuperscript{1},
and Chao Zhang\textsuperscript{1,\textdagger}
}
\IEEEauthorblockA{
\textsuperscript{1}Tsinghua University, China \\
\textsuperscript{2}Huawei Technologies Co., Ltd, China \\
\textsuperscript{*}Equal contribution \qquad
\textsuperscript{\textdagger}Corresponding author \\
}
}

\maketitle

\begin{abstract}
In bandwidth-constrained communication such as satellite and underwater channels, speech must often be transmitted at ultra-low bitrates where intelligibility is the primary objective.
At such extreme compression levels, codecs trained with acoustic reconstruction losses tend to allocate bits to perceptual detail, leading to substantial degradation in word error rate (WER).
This paper proposes ClariCodec, a neural speech codec operating at 300 bits per second (bps) that reformulates quantisation as a stochastic policy, enabling reinforcement learning (RL)-based optimisation of intelligibility.
Specifically, the encoder is fine-tuned using WER-driven rewards while the acoustic reconstruction pipeline remains frozen.
Even without RL, ClariCodec achieves 4.64\% WER on the LibriSpeech test-clean set at 300 bps, already competitive with codecs operating at higher bitrates.
Further RL fine-tuning reduces WER to 3.55\% on test-clean, corresponding to a 23.5\% relative reduction while preserving perceptual quality.
In addition, we adapt ClariCodec to a streaming configuration and show that the proposed RL-based optimisation remains effective under streaming constraints, achieving 4.53\% WER on test-clean with a theoretical latency of 374 ms.

\end{abstract}


\section{Introduction}
\label{sec:1}

In bandwidth-constrained and reliability-limited environments such as satellite and underwater communication, the available transmission capacity may be restricted to only a few hundred bits per second (bps) \cite{Stojanovic2000Underwater, wojcicki2025low}.
Under such conditions, the objective of speech coding shifts from preserving waveform fidelity to ensuring speech intelligibility, where the reliable recovery of linguistic content becomes the primary criterion for success.
This constraint motivates the development of ultra-low-bitrate coding strategies that prioritise semantic clarity while reducing the need to reconstruct fine-grained acoustic details.

Neural speech codecs \cite{zeghidour2021soundstream, defossez2023high} represent audio as compact discrete token sequences for efficient low-bitrate transmission.
Most such studies follow an encoder-quantizer-decoder architecture, where the encoder maps the input waveform to a latent representation, a quantisation module such as residual vector quantisation or finite scalar quantisation (FSQ) \cite{mentzer2024finite} discretises it into tokens, and the decoder reconstructs an approximate waveform. 

However, model structures and training paradigms remain largely rooted in waveform reconstruction objectives.
High-fidelity codecs primarily focus on preserving fine-grained acoustic detail by improving vector quantisation schemes \cite{Kumar2023dac, gu2024esc, Niu2024ndvq, Chae2025vrvq, yang2023hifi, zheng2024srcodec, siuzdak2024snac, guo2024socodec}, model structures \cite{Kumar2023dac, wu2025ts3codec, gu2024esc, yang2024generative, ai2024apcodec, ai2024low, ahn2024hilcodec, zhang2024supercodec, ji2025wavtokenizer, xin2024bigcodec, jiang2023latent, Jenrungrot2023lmcodec, Chae2025vrvq, Bie2025sdcodec, Casanova2025lfsc, liu2024semanticodec, li2025flexicodec, parker2025stablecodec, guo2024socodec, siahkoohi2022ultra, pan2024promptcodec}, or explicit feature disentanglement \cite{guo2024lscodec, li2024single, jiang2023disentangled, ren2024ticodec, zheng2024freecodec, pan2024promptcodec}. 
A parallel line of work has explored semantic codecs, which aim to preserve linguistic content by leveraging representations derived from self-supervised learning \cite{guo2024lscodec, zheng2024freecodec, liu2024semanticodec, zhang2024speechtokenizer, defossez2024moshi, parker2025stablecodec, guo2024socodec, siahkoohi2022ultra, ma2026high, ye2025codec, ye2025llasa, shi2024mmm, mousavi2024should}, automatic speech recognition (ASR) \cite{chen2025sac, li2025flexicodec} or language models \cite{ma2026high, yang2024uniaudio, wang2026towards}. 
Despite these advances, both paradigms face increasing limitations as the bitrate approaches the few-hundred-bps condition. 
In this extreme compression setting, the allocation of bits often becomes misaligned with the information that is most critical for intelligibility, leading to inefficient use of the already scarce transmission budget.

This challenge can be interpreted through the lens of the information bottleneck principle \cite{tishby2000information}. 
Spoken language contains substantial statistical redundancy \cite{denes1963statistics}, which implies that the information required to convey linguistic meaning is far smaller than the full acoustic signal bandwidth \cite{van2017information}. 
In the extreme compression range of around 300 bps, an effective codec must therefore learn representations that remove acoustically redundant detail while retaining the minimal information necessary for intelligible speech recovery. 
Conventional reconstruction objectives, including mel-spectrogram $L_1$ losses and adversarial waveform losses, do not enforce this property.
These losses prioritise acoustic similarity, whereas the automatic evaluation of intelligibility commonly uses WER, which is a discrete and non-differentiable metric. 
As a result, current training paradigms struggle to align bitrate allocation with the information most critical for linguistic decoding.

Beyond bitrate and intelligibility, practical speech communication systems must also operate with low latency.
Many existing neural speech codecs rely on substantial future context, preventing streaming operation and introducing unacceptable latency in interactive communication.
Although several studies have developed causal or streaming codecs \cite{Jenrungrot2023lmcodec,wu2025ts3codec,defossez2024moshi}, preserving intelligibility under both streaming constraints and extreme compression remains underexplored.

To this end, we propose ClariCodec, a neural speech codec designed for extreme compression at 300 bps, and further develop a streaming variant for low-latency operation.\footnote{Audio samples and source code are publicly available at \url{https://demo941.github.io/claricodec-demo-page/} and \url{https://github.com/demo941/ClariCodec}, respectively.}
Both variants utilise a two-stage training strategy: an initial reconstruction-based pre-training phase employing improved FSQ \cite{lin2026ifsq} to establish a stable discrete representation, followed by an RL fine-tuning stage.
In the second stage, we introduce stochastic FSQ, which reformulates the deterministic quantisation grid as a stochastic policy through distance-based probabilistic mapping.
Using group relative policy optimisation (GRPO) \cite{shao2024deepseekmath}, ClariCodec enables direct optimisation against non-differentiable WER reward within a frozen acoustic pipeline.
This approach achieves superior semantic alignment without sacrificing perceptual quality.
For streaming operation, ClariCodec uses causal, state-cached modules together with limited latent look-ahead, enabling streaming encoding and decoding with bounded algorithmic delay.
Experiments on LibriSpeech \cite{panayotov2015librispeech} demonstrate that RL-based optimisation yields a consistent $\sim$23\% relative WER reduction on test-clean, outperforming codecs operating at substantially higher bitrates in terms of intelligibility.
The streaming variant also benefits from RL fine-tuning, reducing WER from 6.63\% to 4.53\% on test-clean with a theoretical latency of 374 ms.
The major contributions of this paper are summarised as follows. 
\begin{itemize}
    \item We propose ClariCodec, a neural speech codec operating at 300 bps that maintains competitive acoustic quality and intelligibility under extreme compression.
    \item We reformulate discrete codec quantisation as a stochastic policy and apply GRPO with a WER-based reward, enabling RL to directly optimise intelligibility. To the best of our knowledge, this is the first study to apply RL for training neural speech codecs.
    \item We extend ClariCodec to streaming operation and demonstrate that the proposed optimisation remains effective under restricted temporal context.
\end{itemize}

\begin{figure*}[ht]
  \centering
  \includegraphics[width=\linewidth]{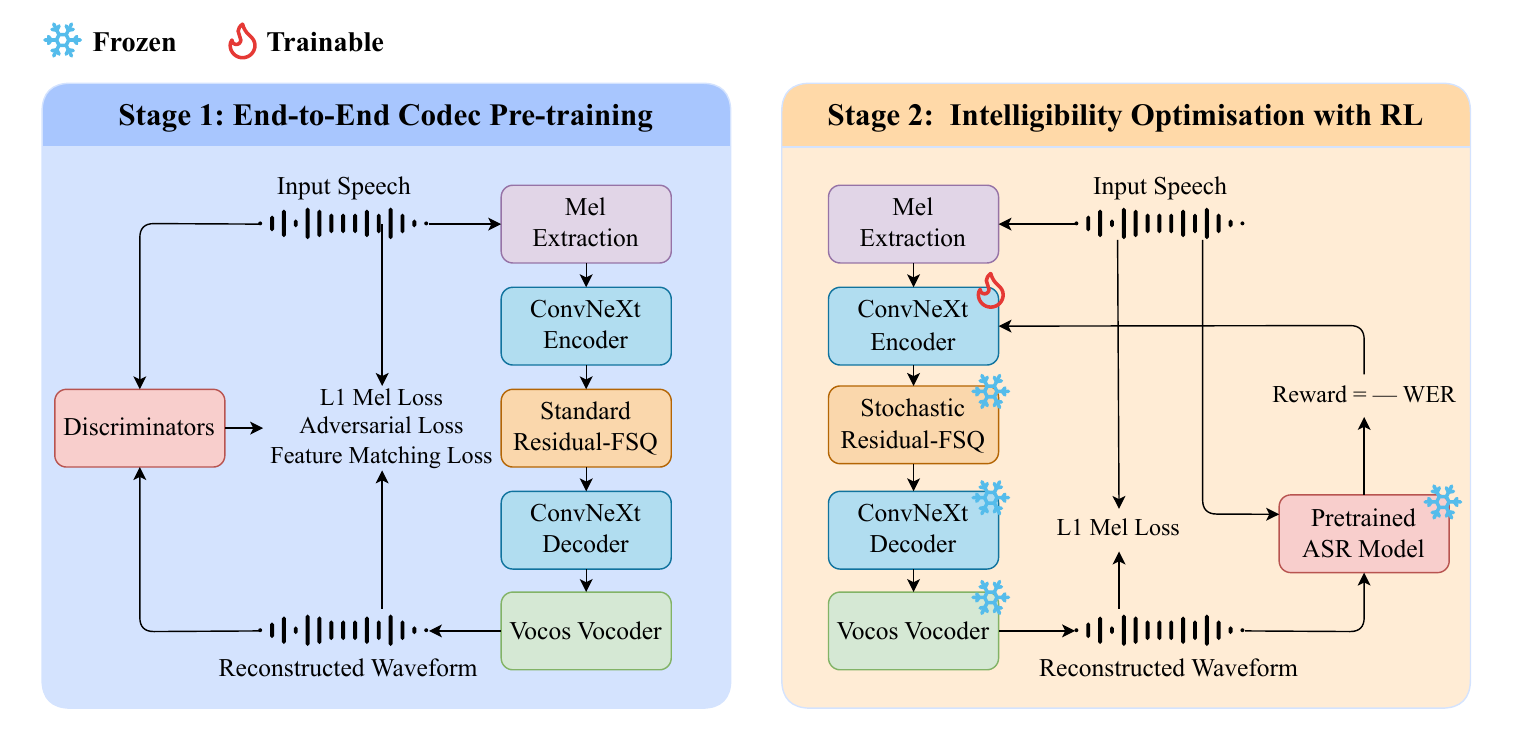}
  \caption{Overview of the two-stage training framework of ClariCodec. In Stage 1, the full codec is trained end-to-end using a combination of $L_1$ mel reconstruction loss, adversarial loss, and feature matching loss to ensure high-fidelity speech reconstruction. In Stage 2, all modules except the encoder are frozen, and the encoder is fine-tuned using an RL objective where the reward signal is derived from a pretrained ASR model, explicitly optimising for speech intelligibility. An $L_1$ mel reconstruction loss is used for preventing perceptual degradation during RL optimisation.}
  \label{fig:overall}
\end{figure*}

\section{Method}
\label{sec:2}

Achieving intelligible speech compression at 300 bps requires balancing acoustic fidelity against the preservation of semantic information, a trade-off that standard reconstruction objectives fail to address effectively.
This work tackles this challenge through a two-stage training strategy that explicitly optimises for semantic information.

\subsection{Model Architecture}
\label{sec:2.1}
\subsubsection{Overview}
\label{sec:2.1.1}

As illustrated in Fig.~\ref{fig:overall}, the model operates on log-mel spectrograms extracted with a hop size of 160 samples (10 ms).
A ConvNeXt V2-based encoder \cite{woo2023convnext} compresses the input into discrete codec indices, and a symmetric decoder reconstructs the log-mel spectrogram from the index sequence.
The reconstructed spectrogram is then converted to waveform samples by a Vocos vocoder \cite{siuzdak2023vocos} trained from scratch jointly with the codec.

To achieve 300 bps, the encoder applies a total temporal downsampling factor of $8\times$ via three successive $2\times$ layers interleaved with ConvNeXt V2 blocks, each halving the temporal resolution while doubling the channel dimension, yielding a latent frame rate of 12.5 Hz. The decoder mirrors this with three $2\times$ upsampling blocks.
Each resampling block adds a fixed average-pooling or nearest-neighbour-interpolation shortcut to a learnable convolutional branch for downsampling or upsampling, respectively.

\subsubsection{Streaming Adaptation}
\label{sec:2.1.2}

For streaming operation, a sliding-window feature extractor buffers the waveform and emits newly available log-mel frames as audio chunks arrive.
All temporal convolutions in the encoder, decoder, and Vocos vocoder are replaced with causal convolutions that cache the historical context required for processing subsequent chunks.
For transposed convolutions, partial outputs within the overlapping region are retained and combined with the outputs produced from the next chunk.
Similarly, the iSTFT module maintains its overlap-add state across chunks to ensure continuous waveform reconstruction.
Together, these stateful operations enable chunk-wise processing without repeatedly recomputing past inputs.

Enforcing strict causality throughout the model, however, restricts the available temporal context and can degrade reconstruction quality.
To achieve a more favourable trade-off between reconstruction quality and latency, we introduce a finite-look-ahead convolution at the encoder output.
This module allows each latent representation to incorporate three future codec frames, while keeping the remaining operations causal.
Given the causal encoding pipeline and the finite future context, the theoretical latency is
\begin{equation}
D_{\mathrm{theory}}
=
\frac{N_{\mathrm{FFT}}+(r-1+N_{\mathrm{la}}r)H}{f_s},
\end{equation}
where $N_{\mathrm{FFT}}=1024$ is the analysis window size, $H=160$ is the hop size, $r=8$ is the temporal downsampling ratio, $N_{\mathrm{la}}=3$ is the number of look-ahead codec frames, and $f_s=16$ kHz is the sampling rate.
This configuration results in a theoretical latency of 374 ms.
This theoretical latency denotes the minimum duration of input audio that must be observed before valid output audio can be produced, excluding computation time on the deployment device.

\subsubsection{Stochastic Residual Quantization}
\label{sec:2.1.3}

To constrain the transmission bandwidth to 300 bps, a residual FSQ (R-FSQ) module with two residual layers is employed. 
Each layer is configured with level dimensions $\mathcal{L} = [8, 8, 8, 8]$, corresponding to an effective codebook size of 12 bits per layer. 
Given a latent frame rate of 12.5 Hz, the overall bitrate is fixed at
\begin{equation}
\mathrm{Bitrate} = 12.5 \mathrm{ Hz} \times 24 \mathrm{ bits/frame} = 300 \mathrm{ bps}.
\end{equation}

To stabilise quantisation, improved FSQ (iFSQ) \cite{lin2026ifsq} replaces the conventional hyperbolic tangent bounding function with a sigmoid activation designed to better match the latent distribution, thereby maximising codebook utilisation.

A central design choice is to reformulate quantisation as stochastic sampling, enabling the encoder to serve as a trainable policy $\pi_\theta$ for RL optimisation in Stage 2.
Instead of deterministically rounding to the nearest level, the negative squared distances to each grid point are treated as logits, and the quantisation level $k_d$ is sampled via Gumbel-Softmax \cite{jang2016categorical}:
\begin{equation}
    \pi(k_d \mid z_d) = \mathrm{Softmax}\left( \frac{-\alpha(z_d - g_{k})^2}{\tau} + \gamma_k \right),
\end{equation}
where $g_k \in \mathcal{G}_d$ denotes the $k$-th grid level, $\alpha$ is the distance scaling factor, $\gamma_k \sim \mathrm{Gumbel}(0,1)$ represents Gumbel noise which is independently sampled for each grid level, and $\tau$ is the temperature parameter.
This stochastic formulation renders the quantiser a differentiable policy $\pi_{\theta}$, which can subsequently be optimised using policy gradient-based approaches.

\subsection{Stage 1: Reconstruction-based Pre-training}
\label{sec:2.2}
The model is optimised end-to-end to minimise a composite loss function:
\begin{equation}
\mathcal{L}_\mathrm{G} = \lambda_\text{rec}\mathcal{L}_\text{rec} + \lambda_\text{adv}\mathcal{L}_\text{adv} + \lambda_\text{fm}\mathcal{L}_\text{fm},
\end{equation}
where $\mathcal{L}_\text{rec}$, $\mathcal{L}_\text{adv}$ and $\mathcal{L}_\text{fm}$ denote the reconstruction, adversarial and feature matching objectives, with $\lambda_\text{rec}$, $\lambda_\text{adv}$, and $\lambda_\text{fm}$ as their respective weights.
The reconstruction loss minimises the $L_1$ distance between log-mel spectrograms of reconstructed and ground-truth audio.
The adversarial loss adopts a Hinge GAN \cite{miyato2018spectral} objective with an ensemble of three discriminators following the Vocos \cite{siuzdak2023vocos} framework, a multi-period discriminator ($\mathcal{L}_\text{adv\text{-}mpd}$) \cite{kong2020hifi} on raw waveforms, a multi-resolution discriminator ($\mathcal{L}_\text{adv\text{-}mrd}$) \cite{jang2021univnet} on complex STFT representations, and a multi-scale discriminator ($\mathcal{L}_\text{adv\text{-}msd}$) \cite{kumar2019melgan} on log-mel spectrograms:
\begin{equation}
\mathcal{L}_\text{adv} = \mathcal{L}_\text{adv-msd} + \mathcal{L}_\text{adv-mpd} + \lambda_\text{mrd}\mathcal{L}_\text{adv-mrd},
\end{equation}
where $\lambda_{\text{mrd}}$ balances the loss magnitudes across discriminators.
The feature matching loss $\mathcal{L}_\text{fm}$ minimises the $L_1$ distance between intermediate discriminator representations of ground-truth and reconstructed samples:
\begin{equation}
\mathcal{L}_\text{fm} = \mathcal{L}_\text{fm-msd} + \mathcal{L}_\text{fm-mpd} + \lambda_\text{mrd}\mathcal{L}_\text{fm-mrd}.
\end{equation}

\subsection{Stage 2: RL-Driven Semantic Optimisation}
\label{sec:2.3}

To maintain the acoustic quality established in Stage 1, all parameters of the quantiser, decoder, and vocoder are frozen during Stage 2, thereby fixing the mapping from discrete tokens to waveform.
The encoder is then modelled as a stochastic policy $\pi_\theta$ over quantisation actions, following the formulation in Section~\ref{sec:2.1.3}, rather than producing deterministic indices.
This formulation recasts discrete quantisation as a differentiable decision process, enabling the encoder to explore semantically improved token configurations within the fixed acoustic space.
Both the reconstructed and ground-truth waveforms are transcribed by a pre-trained ASR system, and the negative WER between the two transcriptions serves as the reward signal.Task-level rewards have also been used in RL-based speech enhancement, including sound-quality and recognition measures \cite{koizumi2017dnn,shen2019rl,kumar2025rlhf}.

The GRPO framework \cite{shao2024deepseekmath} is adopted to optimise the stochastic quantizer. For each input $x$, a group of $G$ token sequences $\{o_i\}_{i=1}^G$ is sampled from the stochastic quantization distribution. Each sampled sequence $o_i$ consists of $L$ discrete codec tokens $\{o_i^{(1)}, o_i^{(2)}, \dots, o_i^{(L)}\}$. 

Given the corresponding sequence-level rewards $\{R_i\}_{i=1}^G$, the group-relative advantage for the $i$-th sample is computed as:
\begin{equation}
\hat{A}_i=
\frac{
R_i-\operatorname{mean}(\{R_i\}_{i=1}^G)
}{
\operatorname{stddev}(\{R_i\}_{i=1}^G)
}.
\end{equation}

Here, $\operatorname{mean}(\cdot)$ and $\operatorname{stddev}(\cdot)$ denote the mean and standard deviation operators, respectively.

Optimisation solely through RL risks degrading perceptual quality by sacrificing speaker fidelity for intelligibility gains.
However, Stage 1 is trained with deterministic quantisation and therefore does not yield a stochastic reference policy.
Although a categorical distribution can be constructed from the Stage 1 model using the distance-based sampling formulation, its probability mass is determined by the introduced stochastic parameterisation and temperature rather than learned during pre-training, making it an unsuitable reference for KL regularisation.
We therefore incorporate a mel spectrogram reconstruction loss to anchor the policy to the original acoustic characteristics.
The final loss is:
\begin{equation}
\begin{split}
L_{\text{total}} = &-\lambda_\text{RL} \mathbb{E}_{{\{o_i\}}_{i=1}^G\sim\pi_\theta} \\&\bigg[\sum_{i=1}^{G}
\bigg(\sum_{l=1}^{L}\log \pi_\theta(o_{i}^{(l)}|x)\bigg)\hat{A}_i\bigg] + \lambda_\text{mel} \mathcal{L}_{\text{mel}},
\end{split}
\end{equation}
where $\lambda_\text{RL}$ and $\lambda_\text{mel}$ denote the weights for the RL and mel reconstruction terms, respectively.

\begin{table*}[ht]
  \caption{Performance comparison of ClariCodec and baseline neural speech codecs on LibriSpeech test-clean. ``\#Param'' denotes the reported number of codec model parameters, and ``\#hours'' denotes the amount of audio directly used for codec training. ``External prior'' indicates whether an externally pre-trained model is used as a codec component, training signal, or reward model. Acoustic quality is evaluated using STOI, PESQ, MOS, UTMOS \cite{saeki2022utmos}, and SIM, while speech intelligibility is measured by WER. Bold indicates the best result in each column. Despite operating at the lowest bitrate, ClariCodec maintains competitive acoustic quality and intelligibility, with further improvements in intelligibility after RL fine-tuning.}
  \label{tab:main}
  \centering
  \small
  \setlength{\tabcolsep}{3.3pt}
  \begin{tabular}{l c c c c c c c c c c c}
    \toprule
    \multirow{2}{*}{Model} & \multirow{2}{*}{\# Param} & \multirow{2}{*}{\# hours} & \multirow{2}{*}{External prior} & \multirow{2}{*}{frame rate} & \multirow{2}{*}{bps} 
    & \multicolumn{6}{c}{test-clean} \\
    \cmidrule{7-12}
    & & & & & 
    & STOI $\uparrow$ & PESQ $\uparrow$ & MOS $\uparrow$ & UTMOS $\uparrow$ & SIM $\uparrow$ & WER(\%) $\downarrow$\\
    \midrule
    Ground Truth & - & - & - & - & -
    & 1.00 & 4.64 & 4.19 & 4.09 & 1.00 & 1.50 \\
    \midrule
    Encodec\cite{defossez2023high} & 15M & 17.5k & No & 75 & 750 
    & 0.77 & 1.25 & 1.10 & 1.25 & 0.25 & 16.1  \\

    StableCodec-700\cite{parker2025stablecodec} & 950M & 105k & Yes & 25 & 700 
    & 0.89 & 1.92 & \textbf{4.52} & \textbf{4.31} & 0.58 & 3.91 \\

    FlexiCodec\cite{li2025flexicodec} & 450M & 54k & Yes & 6.25 & 640
    & \textbf{0.90} & \textbf{2.20} & 3.99 & 4.15 & 0.71 & 2.57 \\

    SAC\cite{chen2025sac} & 533M & 20k & Yes & 12.5/25 & 525
    & \textbf{0.90} & 2.16 & 4.31 & 4.27 & \textbf{0.78} & \textbf{2.00}  \\

    WavTokenizer\cite{ji2025wavtokenizer} & 72M & 8k & No & 40 & 480
    & 0.85 & 1.63 & 3.56 & 3.57 & 0.51 & 7.38  \\

    SoCodec\cite{guo2024socodec} & 54M & 7.2k & Yes & 8.3 & 466
    & 0.09 & 1.28 & 2.58 & 2.50 & 0.39 & 5.59  \\

    StableCodec-400\cite{parker2025stablecodec} & 950M & 105k & Yes & 25 & 400
    & 0.89 & 1.92 & 4.41 & \textbf{4.31} & 0.53 & 4.88 \\

    SemantiCodec\cite{liu2024semanticodec} & 507M & 37.6k & Yes & 12.5/12.5 & 312.5
    & 0.78 & 1.38 & 3.02 & 2.72 & 0.34 & 22.7  \\

    \midrule
    ClariCodec (w/o RL) & 301M & 50k & No & 12.5 & \textbf{300}
    & 0.87 & 1.88 & 3.98 & 4.12 & 0.50 & 4.64  \\

    ClariCodec & 301M & 50k & Yes & 12.5 & \textbf{300}
    & 0.87 & 1.87 & 4.06 & 4.16 & 0.50 & 3.55  \\
    \bottomrule
  \end{tabular}
\end{table*}

\section{Experimental Setup}
\label{sec:3}
\subsection{Datasets}
\label{sec:3.1}
The large subset of Libriheavy \cite{kang2024libriheavy}, comprising 50,000 hours of speech, is used for training. 
Evaluation is conducted on the test-clean and test-other subsets of LibriSpeech \cite{panayotov2015librispeech}. 
Out-of-domain generalisation is further evaluated on LJSpeech \cite{ito2017ljspeech}.
All audio is single-channel at 16\,kHz.

\subsection{Metrics and Baselines}
\label{sec:3.2}
System performance is evaluated across speech intelligibility and acoustic quality. 
Intelligibility is measured by short-time objective intelligibility (STOI) \cite{Taal2010stoi} and WER, where WER is computed using a NeMo Conformer-Transducer\footnote{\url{https://huggingface.co/nvidia/stt_en_conformer_transducer_xlarge}} \cite{gulati2020conformer}. 
Acoustic quality is assessed by perceptual evaluation of speech quality (PESQ) \cite{Rix2001pesq}, UTMOS \cite{saeki2022utmos}, and speaker similarity (SIM), where SIM is computed by a WavLM-based \cite{chen2022wavlm} speaker verification model\footnote{\url{https://github.com/microsoft/UniSpeech/tree/main/downstreams/speaker_verification}}. 
Subjective quality is evaluated using MOS listening tests on 20 randomly selected LibriSpeech test-clean utterances, each lasting 5--10 seconds.
For each source utterance, the ground-truth audio and reconstructed samples from all evaluated systems are rated by five different listeners, and system-level MOS is obtained by averaging over all ratings.

Seven baseline systems are selected for comparison: EnCodec \cite{defossez2023high}, StableCodec \cite{parker2025stablecodec}, FlexiCodec \cite{li2025flexicodec}, SAC \cite{chen2025sac}, WavTokenizer \cite{ji2025wavtokenizer}, SoCodec \cite{guo2024socodec}, and SemantiCodec \cite{liu2024semanticodec}. For EnCodec, we use the first level of RVQ to achieve a bitrate of 750 bps. All baselines are evaluated using their respective official checkpoints. 

\subsection{Training Setup}
\label{sec:3.3}
In Stage 1, the model is trained for 500k steps on 8 NVIDIA H200 GPUs with a batch size of 64, with audio randomly cropped to approximately 4 seconds. 
The loss coefficients are set to $\lambda_\text{rec} = 15$, $\lambda_\text{adv} = 1$, $\lambda_\text{fm} = 1$, and $\lambda_\text{mrd} = 0.2$.

In Stage 2, the model is further trained for 100k steps on 8 NVIDIA H200 GPUs with a batch size of 8 and a GRPO group size of 16, where audio inputs are cropped to approximately 5-second segments.
The loss coefficients are set to $\lambda_\text{RL} = 10$ and $\lambda_\text{mel} = 1$. 
WER rewards are computed using a 1.1B-parameter Hybrid FastConformer TDT-CTC model\footnote{\url{https://huggingface.co/nvidia/parakeet-tdt_ctc-1.1b}} \cite{Rekesh2023fastcon}.

Both stages use the AdamW \cite{loshchilov2019decoupled} optimiser with $\beta_1 = 0.8$ and $\beta_2 = 0.9$, with a one-cycle learning rate schedule comprising a cosine warm-up over the first $5\%$ of updates followed by cosine decay. 
The peak learning rate is $1e-3$ in Stage 1, and $1e-5$ in Stage 2.

\section{Experimental Results}
\label{sec:4}

\subsection{Main Results}
\label{sec:4.1}

Table~\ref{tab:main} compares ClariCodec against existing neural speech codecs across a range of bitrates (300--750 bps) on LibriSpeech test-clean.
ClariCodec operates at 300~bps, the lowest bitrate among all evaluated systems, with the primary objective of
improving speech intelligibility under extreme compression.
Despite operating at only 300 bps, ClariCodec achieves 3.55\% WER, outperforming StableCodec-400 (4.88\%) which operates at 1.33$\times$ the bitrate, showing that intelligibility-oriented training can compensate for the bitrate disadvantage.
While FlexiCodec and SAC achieve better PESQ and SIM scores, these systems operate at 640 and 525 bps, respectively, more than 1.75$\times$ the bitrate of ClariCodec. 
ClariCodec achieves a comparable UTMOS of 4.16, maintaining a predicted MOS competitive with these higher-bitrate systems on this metric, while operating at a substantially lower bitrate.
RL fine-tuning yields a consistent intelligibility improvement, reducing WER from 4.64\% to 3.55\% (23.5\% relative) on test-clean. 
The subjective MOS also increases from 3.98 to 4.06, indicating that this intelligibility gain does not come at the cost of perceived quality.

\subsection{Analysis of Streaming Codec Performance}
\label{streaming}
\begin{table}[!ht]
  \caption{Performance of the streaming ClariCodec variant on LibriSpeech test-clean. The streaming model uses a 374 ms theoretical latency and follows the same two-stage training pipeline as the non-streaming codec.}
  \label{tab:streaming}
  \centering
  \setlength{\tabcolsep}{3pt}
  \resizebox{\linewidth}{!}{
  \begin{tabular}{lccccc}
    \toprule
    Stage & STOI $\uparrow$ & PESQ $\uparrow$ & UTMOS $\uparrow$ & SIM $\uparrow$ & WER(\%) $\downarrow$ \\
    \midrule
    Stage 1 & 0.85 & 1.62 & 3.54 & 0.48 & 6.63 \\
    Stage 2 & 0.85 & 1.63 & 3.65 & 0.48 & 4.53 \\
    \bottomrule
  \end{tabular}}
\end{table}

Table~\ref{tab:streaming} reports the performance of a streaming variant built upon ClariCodec.
Compared with the non-streaming model, the streaming codec introduces a 374 ms theoretical latency, making the system more suitable for low-latency communication scenarios while retaining the same two-stage training paradigm.
The Stage 1 streaming baseline achieves 6.63\% WER on LibriSpeech test-clean.
After Stage 2 RL fine-tuning, WER decreases to 4.53\%, corresponding to a 31.7\% relative reduction, while STOI, PESQ, and SIM remain unchanged.
These results indicate that the proposed RL-based optimisation strategy remains effective under streaming constraints, where the codec must operate with restricted temporal context and low latency.

\subsection{Analysis of Stage 2 Training Strategy}
\label{strategy}
\begin{table}[!ht]
  \caption{Ablation study of Stage 2 loss components on LibriSpeech test-clean. ``Stage 1'' denotes the reconstruction-trained baseline without RL fine-tuning. ``only RL loss'' applies policy gradient optimisation without the mel anchor. ``Mel + RL loss'' adds the mel spectrogram reconstruction loss as an acoustic constraint.}
  \label{tab:strategy}
  \centering
  \setlength{\tabcolsep}{2pt}
  \resizebox{\linewidth}{!}{
  \begin{tabular}{lccccc}
    \toprule
    Loss & STOI $\uparrow$ & PESQ $\uparrow$ & UTMOS $\uparrow$ & SIM $\uparrow$ & WER(\%) $\downarrow$ \\
    \midrule
    Stage 1            & 0.87 & 1.88 & 4.12 & 0.50 & 4.64 \\
    \midrule
    only RL loss       & 0.87 & 1.83 & 4.15 & 0.50 & 3.54 \\
    Mel + RL loss       & 0.87 & 1.87 & 4.16 & 0.50 & 3.55 \\
    \bottomrule
  \end{tabular}}
\end{table}

Table~\ref{tab:strategy} presents an ablation study of the Stage 2 loss design.
We compare three configurations: the Stage 1 reconstruction baseline (no RL fine-tuning), RL optimisation alone (only RL loss), and the proposed combination of RL with a mel spectrogram reconstruction regulariser (Mel + RL loss).
Applying RL fine-tuning without any acoustic constraint yields a substantial reduction in WER—from 4.64 to 3.54 ( 23.7\% relative)—confirming that the optimisation effectively steers the model towards more intelligible outputs.
However, this gain comes at the cost of perceptual quality: PESQ drops from 1.88 to 1.83, indicating that unconstrained RL optimisation tends to sacrifice acoustic fidelity in favour of intelligibility under extreme bitrate constraints.
Introducing the mel reconstruction loss as an acoustic regulariser largely recovers this degradation, raising PESQ from 1.83 to 1.87, while preserving nearly all of the intelligibility benefit.
Notably, PESQ under Mel + RL loss does not fully recover to the Stage 1 level, suggesting an inherent trade-off between acoustic fidelity and semantic optimisation under extreme bitrate constraints.
STOI, UTMOS and SIM remain stable across configurations, suggesting that the mel anchor does not interfere with speaker similarity or overall signal integrity.
Together, these results validate the design choice of combining RL fine-tuning with a mel reconstruction constraint, achieving a favourable balance between intelligibility improvement and acoustic quality preservation.

\subsection{Analysis of Reward Robustness}
\label{reward}
\begin{table}[!ht]
  \caption{Analysis of Stage 2 reward design on LibriSpeech test-clean. The upper block compares different reward recognisers under the WER reward, while the lower block compares different error-rate reward granularities.}
  \label{tab:reward}
  \centering
  \setlength{\tabcolsep}{2pt}
  \resizebox{\linewidth}{!}{
  \begin{tabular}{llccccc}
    \toprule
    Reward model & Reward & STOI $\uparrow$ & PESQ $\uparrow$ & UTMOS $\uparrow$ & SIM $\uparrow$ & WER(\%) $\downarrow$ \\
    \midrule
    \multicolumn{2}{l}{Stage 1} & 0.87 & 1.88 & 4.12 & 0.50 & 4.64 \\
    \midrule
    A & WER & 0.87 & 1.87 & 4.16 & 0.50 & 3.55 \\
    B & WER & 0.87 & 1.89 & 4.15 & 0.50 & 3.59 \\
    C & WER & 0.87 & 1.88 & 4.16 & 0.50 & 3.52 \\
    \midrule
    A & CER & 0.87 & 1.89 & 4.15 & 0.50 & 3.58 \\
    D & PER & 0.87 & 1.79 & 4.20 & 0.49 & 3.51 \\
    \bottomrule
  \end{tabular}}
\end{table}

Table~\ref{tab:reward} analyses the robustness of the Stage 2 reward design from two perspectives: the recogniser used to compute the reward and the granularity of the error-rate objective.
In this table, Model A is the Parakeet-TDT-CTC recogniser\footnote{\url{https://huggingface.co/nvidia/parakeet-tdt_ctc-1.1b}} used in our main experiments, Models B\footnote{\url{https://huggingface.co/facebook/wav2vec2-large-960h-lv60-self}} and C\footnote{\url{https://huggingface.co/nvidia/stt_en_conformer_transducer_xlarge}} are alternative ASR recognisers used for WER rewards, and Model D\footnote{\url{https://huggingface.co/facebook/wav2vec2-lv-60-espeak-cv-ft}} is a phoneme recogniser used only for PER rewards.

When WER is used as the reward, replacing the reward ASR model with different pretrained recognisers still yields consistent WER reductions over the Stage 1 baseline.
This indicates that the proposed codec RL framework is not tightly coupled to a particular recogniser.

We further compare the WER setting with character error rate (CER) and phoneme error rate (PER) rewards.
Although all three reward formulations improve intelligibility over the Stage 1 codec, their final WERs are comparable, with PER yielding the lowest value in this experiment.
These results suggest that the proposed RL framework is not highly sensitive to either the reward recogniser or the granularity of the error-rate objective.

\subsection{Robustness under Challenging Speech Conditions}
\label{test_other}
\begin{table}[!ht]
  \caption{Evaluation on LibriSpeech test-other, a challenging benchmark used to assess robustness under challenging speech conditions.}
  \label{tab:test_other}
  \centering
  \setlength{\tabcolsep}{2pt}
  \resizebox{\linewidth}{!}{
  \begin{tabular}{lccccc}
    \toprule
    Model & STOI $\uparrow$ & PESQ $\uparrow$ & UTMOS $\uparrow$ & SIM $\uparrow$ & WER(\%) $\downarrow$ \\
    \midrule
    Ground Truth     
    & 1.00 & 4.64 & 3.50 & 1.00 & 2.81 \\
    \midrule
    Encodec\cite{defossez2023high} 
    & 0.76 & 1.27 & 1.26 & 0.25 & 36.4 \\

    StableCodec-700\cite{parker2025stablecodec}
    & 0.87 & 1.91 & 3.91 & 0.58 & 12.0 \\

    FlexiCodec\cite{li2025flexicodec}
    & \textbf{0.88} & \textbf{2.11} & 3.74 & 0.71 & 4.69 \\

    SAC\cite{chen2025sac} 
    & 0.87 & 2.04 & 3.90 & \textbf{0.77} & \textbf{4.15} \\

    WavTokenizer\cite{ji2025wavtokenizer} 
    & 0.82 & 1.60 & 3.16 & 0.48 & 21.1 \\

    SoCodec\cite{guo2024socodec} 
    & 0.09 & 1.33 & 2.33 & 0.46 & 10.6 \\

    StableCodec-400\cite{parker2025stablecodec} 
    & 0.85 & 1.79 & \textbf{3.93} & 0.53 & 14.4 \\

    SemantiCodec\cite{liu2024semanticodec} 
    & 0.76 & 1.39 & 2.41 & 0.38 & 40.2 \\

    \midrule
    ClariCodec (w/o RL) 
    & 0.84 & 1.75 & 3.67 & 0.51 & 13.3 \\

    ClariCodec 
    & 0.84 & 1.75 & 3.73 & 0.51 & 10.4 \\
    \bottomrule
  \end{tabular}}
\end{table}

Table~\ref{tab:test_other} reports the results on LibriSpeech test-other, which contains substantially more challenging utterances than test-clean.
Overall, the trends are largely consistent with those observed on test-clean, with all codecs experiencing a performance degradation due to the increased difficulty of the dataset.
Importantly, the benefits of RL fine-tuning generalise beyond clean speech.
Compared with the reconstruction-only model, RL optimisation reduces WER from 13.3\% to 10.4\% (21.8\% relative) while leaving STOI, PESQ, and speaker similarity unchanged.

\subsection{Out-of-Domain Generalisation}
\label{ljspeech}
\begin{table}[!ht]
  \caption{Out-of-domain evaluation on LJSpeech to assess cross-speaker and cross-domain generalisation.}
  \label{tab:ljspeech}
  \centering
  \setlength{\tabcolsep}{2pt}
  \resizebox{\linewidth}{!}{
  \begin{tabular}{lccccc}
    \toprule
    Model & STOI $\uparrow$ & PESQ $\uparrow$ & UTMOS $\uparrow$ & SIM $\uparrow$ & WER(\%) $\downarrow$ \\
    \midrule
    Ground Truth     
    & 1.00 & 4.64 & 4.38 & 1.00 & 2.82 \\
    \midrule
    Encodec\cite{defossez2023high} 
    & 0.77 & 1.23 & 1.24 & 0.27 & 18.0 \\

    StableCodec-700\cite{parker2025stablecodec}
    & \textbf{0.95} & \textbf{2.59} & \textbf{4.46} & \textbf{0.91} & 3.51 \\

    FlexiCodec\cite{li2025flexicodec}
    & 0.92 & 2.18 & 4.33 & 0.70 & 3.51 \\

    SAC\cite{chen2025sac} 
    & 0.92 & 2.24 & 4.43 & 0.84 & \textbf{3.12} \\

    WavTokenizer\cite{ji2025wavtokenizer} 
    & 0.87 & 1.72 & 3.69 & 0.48 & 6.93 \\

    SoCodec\cite{guo2024socodec} 
    & 0.14 & 1.27 & 2.53 & 0.39 & 4.51 \\

    StableCodec-400\cite{parker2025stablecodec} 
    & 0.93 & 2.40 & \textbf{4.46} & 0.90 & 3.90 \\

    SemantiCodec\cite{liu2024semanticodec} 
    & 0.82 & 1.43 & 2.65 & 0.38 & 17.4 \\

    \midrule
    ClariCodec (w/o RL) 
    & 0.92 & 2.21 & 4.37 & 0.78 & 4.44 \\

    ClariCodec 
    & 0.92 & 2.21 & 4.38 & 0.77 & 3.82 \\
    \bottomrule
  \end{tabular}}
\end{table}

Table~\ref{tab:ljspeech} evaluates the out-of-domain generalisation capability of ClariCodec on LJSpeech, which differs substantially from LibriSpeech in speaker characteristics, recording conditions, and corpus composition.
Compared with the relatively consistent trends observed on test-clean and test-other, codec performance varies more noticeably on LJSpeech, revealing different levels of robustness to domain shift across different codecs.
RL optimisation reduces WER from 4.44\% to 3.82\% (14.0\% relative) while maintaining nearly unchanged acoustic quality metrics. 

Moreover, the performance gap between ClariCodec and higher-bitrate semantic codecs becomes smaller on LJSpeech. For example, the WER gap relative to SAC decreases from 1.55\% to 0.70\%, while the gap relative to FlexiCodec decreases from 0.98\% to 0.31\%. These results indicate that the proposed representation generalises effectively across domains and that the benefits of WER-guided optimisation extend beyond the training distribution.

\section{Conclusions}
Maintaining speech intelligibility at ultra-low bitrates remains a fundamental challenge for neural speech codecs in bandwidth-constrained environments.
To address this, we presented ClariCodec, a neural speech codec operating at 300 bps that incorporates reinforcement learning to explicitly optimise semantic retention.
By reformulating quantisation as a stochastic policy and leveraging WER-based reward signals, ClariCodec achieves a WER of 3.55\% on LibriSpeech test-clean, outperforming baseline models operating at substantially higher bitrates.
The streaming variant achieves a WER of 4.53\% with a theoretical latency of 374 ms, demonstrating that the proposed optimisation remains effective under streaming constraints.
Future work will further evaluate the effects of operating at such low bitrates on downstream generative tasks, including speech synthesis and codec-based speech large language models, and explore more comprehensive RL objectives that incorporate both WER and acoustic quality metrics as reward signals.
% Future work will focus on two primary directions.
% First, the effects of operating at such low bitrates will be further evaluated on downstream generative tasks, including speech synthesis and codec-based speech large language models.
% Finally, we aim to explore more comprehensive optimisation objectives for RL training, incorporating not only WER but also acoustic quality metrics as reward signals.

\section*{Acknowledgment}
GLM was used for language polishing in the Results section. 

\bibliographystyle{IEEEtran}
\bibliography{mybib}

% Generated by IEEEtran.bst, version: 1.14 (2015/08/26)
\begin{thebibliography}{10}
\providecommand{\url}[1]{#1}
\csname url@samestyle\endcsname
\providecommand{\newblock}{\relax}
\providecommand{\bibinfo}[2]{#2}
\providecommand{\BIBentrySTDinterwordspacing}{\spaceskip=0pt\relax}
\providecommand{\BIBentryALTinterwordstretchfactor}{4}
\providecommand{\BIBentryALTinterwordspacing}{\spaceskip=\fontdimen2\font plus
\BIBentryALTinterwordstretchfactor\fontdimen3\font minus \fontdimen4\font\relax}
\providecommand{\BIBforeignlanguage}[2]{{%
\expandafter\ifx\csname l@#1\endcsname\relax
\typeout{** WARNING: IEEEtran.bst: No hyphenation pattern has been}%
\typeout{** loaded for the language `#1'. Using the pattern for}%
\typeout{** the default language instead.}%
\else
\language=\csname l@#1\endcsname
\fi
#2}}
\providecommand{\BIBdecl}{\relax}
\BIBdecl

\bibitem{Stojanovic2000Underwater}
M.~Stojanovic and J.~Preisig, ``Underwater acoustic communication channels: {Propagation} models and statistical characterization,'' \emph{IEEE Communications Magazine}, vol.~47, no.~1, pp. 84--89, 2009.

\bibitem{wojcicki2025low}
K.~Wojcicki, Y.~Z. Isik, L.~Lechler, M.~Yesilbursa, I.~Balić, W.~Mack, R.~Łaganowski, G.~Zhang, Y.~Adi, M.~Kim, and S.~Watanabe, ``Low-resource audio codec ({LRAC}): 2025 challenge description,'' \emph{arXiv preprint arXiv:2510.23312}, 2025.

\bibitem{zeghidour2021soundstream}
N.~Zeghidour, A.~Luebs, A.~Omran, J.~Skoglund, and M.~Tagliasacchi, ``{SoundStream}: {A}n end-to-end neural audio codec,'' \emph{IEEE/ACM Transactions on Audio, Speech, and Language Processing}, vol.~30, pp. 495--507, 2022.

\bibitem{defossez2023high}
A.~D{\'e}fossez, J.~Copet, G.~Synnaeve, and Y.~Adi, ``High fidelity neural audio compression,'' \emph{Transactions on Machine Learning Research}, 2023.

\bibitem{mentzer2024finite}
F.~Mentzer, D.~Minnen, E.~Agustsson, and M.~Tschannen, ``Finite scalar quantization: {Vq}-vae made simple,'' in \emph{Proc. ICLR}, Vienna, 2024.

\bibitem{Kumar2023dac}
R.~Kumar, P.~Seetharaman, A.~Luebs, I.~Kumar, and K.~Kumar, ``High-fidelity audio compression with improved {RVQGAN},'' in \emph{Proc. NeurIPS}, New Orleans, 2023.

\bibitem{gu2024esc}
Y.~Gu and E.~Diao, ``{ESC}: {Efficient} speech coding with cross-scale residual vector quantized {T}ransformers,'' in \emph{Proc. EMNLP}, Miami, 2024.

\bibitem{Niu2024ndvq}
Z.~Niu, S.~Chen, L.~Zhou, Z.~Ma, X.~Chen, and S.~Liu, ``{NDVQ}: {Robust} neural audio codec with normal distribution-based vector quantization,'' in \emph{Proc. SLT}, Macao, 2024.

\bibitem{Chae2025vrvq}
Y.~Chae, W.~Choi, Y.~Takida, J.~Koo, Y.~Ikemiya, Z.~Zhong, K.~W. Cheuk, M.~A. Martínez-Ramírez, K.~Lee, W.-H. Liao, and Y.~Mitsufuji, ``{Variable} bitrate residual vector quantization for audio coding,'' in \emph{Proc. ICASSP}, Hyderabad, 2025.

\bibitem{yang2023hifi}
D.~Yang, S.~Liu, R.~Huang, J.~Tian, C.~Weng, and Y.~Zou, ``{HiFi-Codec}: {Group}-residual vector quantization for high fidelity audio codec,'' \emph{arXiv preprint arXiv:2305.02765}, 2023.

\bibitem{zheng2024srcodec}
Y.~Zheng, W.~Tu, L.~Xiao, and X.~Xu, ``{Srcodec}: {Split}-residual vector quantization for neural speech codec,'' in \emph{Proc. ICASSP}, Seoul, 2024.

\bibitem{siuzdak2024snac}
H.~Siuzdak, F.~Gr{\"o}tschla, and L.~A. Lanzend{\"o}rfer, ``{SNAC}: Multi-scale neural audio codec,'' in \emph{Audio Imagination: NeurIPS 2024 Workshop AI-Driven Speech, Music, and Sound Generation}, Vancouver, 2024.

\bibitem{guo2024socodec}
H.~Guo, F.~Xie, K.~Xie, D.~Yang, D.~Guo, X.~Wu, and H.~Meng, ``{SoCodec}: {A} semantic-ordered multi-stream speech codec for efficient language model based text-to-speech synthesis,'' in \emph{Proc. SLT}, Macao, 2024.

\bibitem{wu2025ts3codec}
H.~Wu, N.~Kanda, S.~E. Eskimez, and J.~Li, ``{TS3-Codec}: {Transformer}-based simple streaming single codec,'' in \emph{Proc. Interspeech}, Rotterdam, 2025.

\bibitem{yang2024generative}
H.~Yang, I.~Jang, and M.~Kim, ``Generative de-quantization for neural speech codec via latent diffusion,'' in \emph{Proc. ICASSP}, Seoul, 2024.

\bibitem{ai2024apcodec}
Y.~Ai, X.-H. Jiang, Y.-X. Lu, H.-P. Du, and Z.-H. Ling, ``{APCodec}: {A} neural audio codec with parallel amplitude and phase spectrum encoding and decoding,'' \emph{IEEE/ACM Transactions on Audio, Speech, and Language Processing}, vol.~32, pp. 3256--3269, 2024.

\bibitem{ai2024low}
Y.~Ai, Y.-X. Lu, X.-H. Jiang, Z.-Y. Sheng, R.-C. Zheng, and Z.-H. Ling, ``A low-bitrate neural audio codec framework with bandwidth reduction and recovery for high-sampling-rate waveforms,'' in \emph{Proc. Interspeech}, Kos Island, 2024.

\bibitem{ahn2024hilcodec}
S.~Ahn, B.~J. Woo, M.~H. Han, C.~Moon, and N.~S. Kim, ``{HILCodec}: {High}-fidelity and lightweight neural audio codec,'' \emph{IEEE Journal of Selected Topics in Signal Processing}, vol.~18, no.~8, pp. 1517--1530, 2024.

\bibitem{zhang2024supercodec}
Y.~Zheng, W.~Tu, L.~Xiao, and X.~Xu, ``{SuperCodec}: {A} neural speech codec with selective back-projection network,'' in \emph{Proc. ICASSP}, Seoul, 2024.

\bibitem{ji2025wavtokenizer}
S.~Ji, Z.~Jiang, W.~Wang, Y.~Chen, M.~Fang, J.~Zuo, Q.~Yang, X.~Cheng, Z.~Wang, R.~Li, Z.~Zhang, X.~Yang, R.~Huang, Y.~Jiang, Q.~Chen, S.~Zheng, and Z.~Zhao, ``{WavTokenizer}: {An} efficient acoustic discrete codec tokenizer for audio language modeling,'' in \emph{Proc. ICLR}, Singapore, 2025.

\bibitem{xin2024bigcodec}
D.~Xin, X.~Tan, S.~Takamichi, and H.~Saruwatari, ``{BigCodec}: {Pushing} the limits of low-bitrate neural speech codec,'' \emph{arXiv preprint arXiv:2409.05377}, 2024.

\bibitem{jiang2023latent}
X.~Jiang, X.~Peng, H.~Xue, Y.~Zhang, and Y.~Lu, ``{Latent}-domain predictive neural speech coding,'' \emph{IEEE/ACM Transactions on Audio, Speech, and Language Processing}, vol.~31, pp. 2111--2123, 2023.

\bibitem{Jenrungrot2023lmcodec}
T.~Jenrungrot, M.~Chinen, W.~B. Kleijn, J.~Skoglund, Z.~Borsos, N.~Zeghidour, and M.~Tagliasacchi, ``{LMCodec}: {A} low bitrate speech codec with causal {T}ransformer models,'' in \emph{Proc. ICASSP}, Rhodes Island, 2023.

\bibitem{Bie2025sdcodec}
X.~Bie, X.~Liu, and G.~Richard, ``Learning source disentanglement in neural audio codec,'' in \emph{Proc. ICASSP}, Hyderabad, 2025.

\bibitem{Casanova2025lfsc}
E.~Casanova, R.~Langman, P.~Neekhara, S.~Hussain, J.~Li, S.~Ghosh, A.~Jukić, and S.-G. Lee, ``{Low Frame-rate Speech Codec}: {A} codec designed for fast high-quality speech {LLM} training and inference,'' in \emph{Proc. ICASSP}, Hyderabad, 2025.

\bibitem{liu2024semanticodec}
H.~Liu, X.~Xu, Y.~Yuan, M.~Wu, W.~Wang, and M.~D. Plumbley, ``{SemantiCodec}: {An} ultra low bitrate semantic audio codec for general sound,'' \emph{IEEE Journal of Selected Topics in Signal Processing}, vol.~18, no.~8, pp. 1448--1461, 2024.

\bibitem{li2025flexicodec}
J.~Li, Y.~Qian, Y.~Hu, L.~Zhang, X.~Wang, H.~Lu, M.~Thakker, J.~Li, S.~Zhao, and Z.~Wu, ``{FlexiCodec}: {A} dynamic neural audio codec for low frame rates,'' in \emph{Proc. ICLR}, Rio de Janeiro, 2026.

\bibitem{parker2025stablecodec}
J.~D. Parker, A.~Smirnov, J.~Pons, C.~Carr, Z.~Zukowski, Z.~Evans, and X.~Liu, ``Scaling {T}ransformers for low-bitrate high-quality speech coding,'' in \emph{Proc. ICLR}, Singapore, 2025.

\bibitem{siahkoohi2022ultra}
A.~Siahkoohi, M.~Chinen, T.~Denton, W.~B. Kleijn, and J.~Skoglund, ``{Ultra}-low-bitrate speech coding with pretrained {T}ransformers,'' in \emph{Proc. Interspeech}, Incheon, 2022.

\bibitem{pan2024promptcodec}
Y.~Pan, X.~Zhang, Y.~Yang, J.~Yao, Y.~Hu, J.~Ye, H.~Zhou, L.~Ma, and J.~Zhao, ``{PSCodec}: {A} series of high-fidelity low-bitrate neural speech codecs leveraging prompt encoders,'' \emph{arXiv preprint arXiv:2404.02702}, 2024.

\bibitem{guo2024lscodec}
Y.~Guo, Z.~Li, C.~Du, H.~Wang, X.~Chen, and K.~Yu, ``{LSCodec}: {Low}-bitrate and speaker-decoupled discrete speech codec,'' in \emph{Proc. Interspeech}, Rotterdam, 2025.

\bibitem{li2024single}
H.~Li, L.~Xue, H.~Guo, X.~Zhu, Y.~Lv, L.~Xie, Y.~Chen, H.~Yin, and Z.~Li, ``{Single-Codec}: {Single-codebook} speech codec towards high-performance speech generation,'' in \emph{Proc. Interspeech}, Kos Island, 2024.

\bibitem{jiang2023disentangled}
X.~Jiang, X.~Peng, Y.~Zhang, and Y.~Lu, ``Disentangled feature learning for real-time neural speech coding,'' in \emph{Proc. ICASSP}, Rhodes Island, 2023.

\bibitem{ren2024ticodec}
Y.~Ren, T.~Wang, J.~Yi, L.~Xu, J.~Tao, C.~Y. Zhang, and J.~Zhou, ``{Fewer}-token neural speech codec with time-invariant codes,'' in \emph{Proc. ICASSP}, Seoul, 2024.

\bibitem{zheng2024freecodec}
Y.~Zheng, W.~Tu, Y.~Kang, J.~Chen, Y.~Zhang, L.~Xiao, Y.~Yang, and L.~Ma, ``{FreeCodec}: {A} disentangled neural speech codec with fewer tokens,'' in \emph{Proc. Interspeech}, Rotterdam, 2025.

\bibitem{zhang2024speechtokenizer}
X.~Zhang, D.~Zhang, S.~Li, Y.~Zhou, and X.~Qiu, ``{SpeechTokenizer}: {Unified} speech tokenizer for speech language models,'' in \emph{Proc. ICLR}, Vienna, 2024.

\bibitem{defossez2024moshi}
A.~D{\'e}fossez, L.~Mazar{\'e}, M.~Orsini, A.~Royer, P.~P{\'e}rez, H.~J{\'e}gou, E.~Grave, and N.~Zeghidour, ``{Moshi}: {A} speech-text foundation model for real-time dialogue,'' \emph{arXiv preprint arXiv:2410.00037}, 2024.

\bibitem{ma2026high}
H.~Ma, R.~Jing, S.~Liu, C.~Gong, C.~Zhang, X.-L. Zhang, and X.~Li, ``High-fidelity generative audio compression at 0.275 kbps,'' \emph{arXiv preprint arXiv:2602.00648}, 2026.

\bibitem{ye2025codec}
Z.~Ye, P.~Sun, J.~Lei, H.~Lin, X.~Tan, Z.~Dai, Q.~Kong, J.~Chen, J.~Pan, Q.~Liu, Y.~Guo, and W.~Xue, ``{Codec} does matter: {Exploring} the semantic shortcoming of codec for audio language model,'' in \emph{Proc. AAAI}, Philadelphia, 2025.

\bibitem{ye2025llasa}
Z.~Ye, X.~Zhu, C.-M. Chan, X.~Wang, X.~Tan, J.~Lei, Y.~Peng, H.~Liu, Y.~Jin, Z.~Dai, H.~Lin, J.~Chen, X.~Du, L.~Xue, Y.~Chen, Z.~Li, L.~Xie, Q.~Kong, Y.~Guo, and W.~Xue, ``{LLaSa}: {Scaling} train-time and inference-time compute for {LLaMa}-based speech synthesis,'' \emph{arXiv preprint arXiv:2502.04128}, 2025.

\bibitem{shi2024mmm}
J.~Shi, X.~Ma, H.~Inaguma, A.~Sun, and S.~Watanabe, ``{MMM}: {Multi}-layer multi-residual multi-stream discrete speech representation from self-supervised learning model,'' in \emph{Proc. Interspeech}, Kos Island, 2024.

\bibitem{mousavi2024should}
P.~Mousavi, J.~Duret, S.~Zaiem, L.~Della~Libera, A.~Ploujnikov, C.~Subakan, and M.~Ravanelli, ``{How} should we extract discrete audio tokens from self-supervised models?'' in \emph{Proc. Interspeech}, Kos Island, 2024.

\bibitem{chen2025sac}
W.~Chen, X.~Wang, R.~Yan, Y.~Chen, Z.~Niu, Z.~Ma, X.~Li, Y.~Liang, H.~Wen, S.~Yin, M.~Tao, and X.~Chen, ``{SAC}: {Neural} speech codec with semantic-acoustic dual-stream quantization,'' \emph{arXiv preprint arXiv:2510.16841}, 2025.

\bibitem{yang2024uniaudio}
D.~Yang, H.~Guo, Y.~Wang, R.~Huang, X.~Li, X.~Tan, X.~Wu, and H.~Meng, ``{UniAudio} 1.5: {Large} language model-driven audio codec is a few-shot audio task learner,'' in \emph{Proc. NeurIPS}, Vancouver, 2024.

\bibitem{wang2026towards}
S.~Wang, Z.~Jin, C.~Tang, Q.~Li, B.~Li, C.~Chen, Y.~Hu, W.~Yu, Y.~Li, J.~Zhuang \emph{et~al.}, ``Towards general auditory intelligence for machine listening and speaking,'' \emph{Nature Machine Intelligence}, pp. 1--20, 2026.

\bibitem{tishby2000information}
N.~Tishby, F.~C. Pereira, and W.~Bialek, ``The information bottleneck method,'' \emph{arXiv preprint physics/0004057}, 2000.

\bibitem{denes1963statistics}
P.~B. Denes, ``On the statistics of spoken {E}nglish,'' \emph{The Journal of the Acoustical Society of America}, vol.~35, no.~6, pp. 892--904, 1963.

\bibitem{van2017information}
S.~Van~Kuyk, W.~B. Kleijn, and R.~C. Hendriks, ``On the information rate of speech communication,'' in \emph{Proc. ICASSP}, New Orleans, 2017.

\bibitem{lin2026ifsq}
B.~Lin, Z.~Li, Y.~Niu, K.~Gong, Y.~Ge, Y.~Lin, M.~Zheng, J.~Zhang, M.~Yang, Z.~Zhong, L.~Bo, and L.~Yuan, ``{iFSQ}: {Improving} {FSQ} for image generation with 1 line of code,'' \emph{arXiv preprint arXiv:2601.17124}, 2026.

\bibitem{shao2024deepseekmath}
Z.~Shao, P.~Wang, Q.~Zhu, R.~Xu, J.~Song, X.~Bi, H.~Zhang, M.~Zhang, Y.~K. Li, Y.~Wu, and D.~Guo, ``{DeepSeekMath}: {Pushing} the limits of mathematical reasoning in open language models,'' \emph{arXiv preprint arXiv:2402.03300}, 2024.

\bibitem{panayotov2015librispeech}
V.~Panayotov, G.~Chen, D.~Povey, and S.~Khudanpur, ``{LibriSpeech}: An {ASR} corpus based on public domain audio books,'' in \emph{Proc. ICASSP}, Brisbane, 2015.

\bibitem{woo2023convnext}
S.~Woo, S.~Debnath, R.~Hu, X.~Chen, Z.~Liu, I.~S. Kweon, and S.~Xie, ``{ConvNeXt V2}: {Co}-designing and scaling convnets with masked autoencoders,'' in \emph{Proc. CVPR}, Vancouver, 2023.

\bibitem{siuzdak2023vocos}
H.~Siuzdak, ``{Vocos}: {Closing} the gap between time-domain and fourier-based neural vocoders for high-quality audio synthesis,'' in \emph{Proc. ICLR}, Vienna, 2024.

\bibitem{jang2016categorical}
E.~Jang, S.~Gu, and B.~Poole, ``Categorical reparameterization with gumbel-softmax,'' in \emph{Proc. ICLR}, Toulon, 2017.

\bibitem{miyato2018spectral}
T.~Miyato, T.~Kataoka, M.~Koyama, and Y.~Yoshida, ``Spectral normalization for generative adversarial networks,'' in \emph{Proc. ICLR}, Vancouver, 2018.

\bibitem{kong2020hifi}
J.~Kong, J.~Kim, and J.~Bae, ``{HiFi-GAN}: {Generative} adversarial networks for efficient and high fidelity speech synthesis,'' in \emph{Proc. NeurIPS}, 2020, pp. 19\,655--19\,666.

\bibitem{jang2021univnet}
W.~Jang, D.~Lim, J.~Yoon, B.~Kim, and J.~Kim, ``{UnivNet}: {A} neural vocoder with multi-resolution spectrogram discriminators for high-fidelity waveform generation,'' in \emph{Proc. Interspeech}, Brno, 2021.

\bibitem{kumar2019melgan}
K.~Kumar, R.~Kumar, T.~De~Boissiere, L.~Gestin, W.~Z. Teoh, J.~Sotelo, A.~De~Brebisson, Y.~Bengio, and A.~C. Courville, ``{MelGAN}: {Generative} adversarial networks for conditional waveform synthesis,'' in \emph{Proc. NeurIPS}, Vancouver, 2019.

\bibitem{koizumi2017dnn}
Y.~Koizumi, K.~Niwa, Y.~Hioka, K.~Kobayashi, and Y.~Haneda, ``{DNN}-based source enhancement self-optimized by reinforcement learning using sound quality measurements,'' in \emph{Proc. ICASSP}, New Orleans, 2017.

\bibitem{shen2019rl}
Y.-L. Shen, C.-Y. Huang, S.-S. Wang, Y.~Tsao, H.-M. Wang, and T.-S. Chi, ``Reinforcement learning based speech enhancement for robust speech recognition,'' in \emph{Proc. ICASSP}, Brighton, 2019.

\bibitem{kumar2025rlhf}
A.~Kumar, A.~Perrault, and D.~S. Williamson, ``Using {RLHF} to align speech enhancement approaches to mean-opinion quality scores,'' in \emph{Proc. ICASSP}, Hyderabad, 2025.

\bibitem{saeki2022utmos}
T.~Saeki, D.~Xin, W.~Nakata, T.~Koriyama, S.~Takamichi, and H.~Saruwatari, ``{UTMOS}: {UTokyo}-{Sarulab} system for {VoiceMOS} challenge 2022,'' in \emph{Proc. Interspeech}, Incheon, 2022.

\bibitem{kang2024libriheavy}
W.~Kang, X.~Yang, Z.~Yao, F.~Kuang, Y.~Yang, L.~Guo, L.~Lin, and D.~Povey, ``{Libriheavy}: {A} 50,000 hours {ASR} corpus with punctuation casing and context,'' in \emph{Proc. ICASSP}, Seoul, 2024.

\bibitem{ito2017ljspeech}
K.~Ito and L.~Johnson, ``The lj speech dataset,'' \url{https://keithito.com/LJ-Speech-Dataset/}, 2017.

\bibitem{Taal2010stoi}
C.~H. Taal, R.~C. Hendriks, R.~Heusdens, and J.~Jensen, ``A short-time objective intelligibility measure for time-frequency weighted noisy speech,'' in \emph{Proc. ICASSP}, Dallas, 2010.

\bibitem{gulati2020conformer}
A.~Gulati, J.~Qin, C.-C. Chiu, N.~Parmar, Y.~Zhang, J.~Yu, W.~Han, S.~Wang, Z.~Zhang, Y.~Wu, and R.~Pang, ``Conformer: {Convolution}-augmented {T}ransformer for speech recognition,'' in \emph{Proc. Interspeech}, Shanghai, 2020.

\bibitem{Rix2001pesq}
A.~Rix, J.~Beerends, M.~Hollier, and A.~Hekstra, ``Perceptual evaluation of speech quality ({PESQ})-a new method for speech quality assessment of telephone networks and codecs,'' in \emph{Proc. ICASSP}, Salt Lake City, 2001.

\bibitem{chen2022wavlm}
S.~Chen, C.~Wang, Z.~Chen, Y.~Wu, S.~Liu, Z.~Chen, J.~Li, N.~Kanda, T.~Yoshioka, X.~Xiao, J.~Wu, L.~Zhou, S.~Ren, Y.~Qian, Y.~Qian, J.~Wu, M.~Zeng, X.~Yu, and F.~Wei, ``{WavLM}: {Large}-scale self-supervised pre-training for full stack speech processing,'' \emph{{IEEE} Journal of Selected Topics in Signal Processing}, vol.~16, no.~6, pp. 1505--1518, 2022.

\bibitem{Rekesh2023fastcon}
D.~Rekesh, N.~R. Koluguri, S.~Kriman, S.~Majumdar, V.~Noroozi, H.~Huang, O.~Hrinchuk, K.~Puvvada, A.~Kumar, J.~Balam, and B.~Ginsburg, ``Fast conformer with linearly scalable attention for efficient speech recognition,'' in \emph{Proc. ASRU}, Taipei, 2023.

\bibitem{loshchilov2019decoupled}
I.~Loshchilov and F.~Hutter, ``Decoupled weight decay regularization,'' in \emph{Proc. ICLR}, New Orleans, 2019.

\end{thebibliography}

\end{document}